\documentclass[prd,aps,superscriptaddress,nofootinbib,amsmath,amssymb,shownopacs]{revtex4}
\usepackage{graphicx}
\usepackage{dcolumn}
\usepackage{bm}
\usepackage{multirow}
\usepackage{adjustbox}
\usepackage{textcomp}
\usepackage{mathtext}
\usepackage{hyperref}
\graphicspath{{./eps/}}

\begin{document}
\title{\Large $\bar{\nu}_{\mu}$ induced quasielastic production of hyperons leading to pions}
\author{A. \surname{Fatima}}
 \email{atikafatima1706@gmail.com}
\affiliation{Department of Physics, Aligarh Muslim University, Aligarh-202002, India}
\author{M. Sajjad \surname{Athar}}
\affiliation{Department of Physics, Aligarh Muslim University, Aligarh-202002, India}
\author{S. K. \surname{Singh}}
\affiliation{Department of Physics, Aligarh Muslim University, Aligarh-202002, India}

\begin{abstract}
The quasielastic production cross sections and polarizations of the hyperons induced by ${\bar\nu}_\mu$ on the free nucleon as well as from $^{40}$Ar in the sub-GeV energy region has been reviewed~\cite{Fatima:2018wsy,Fatima:2018tzs,Akbar:2016awk,Fatima:2018gjy,Akbar:2017qsf}. 
The effects of the second class currents in the axial vector sector with and without T-invariance as well as the effect
 of SU(3) symmetry breaking are also studied. We find that the cross sections
and the various polarization components can effectively be used to determine the axial
vector transition form factors in the strangeness sector and to test the validity of various
symmetries of the weak hadronic currents like G-invariance, T-invariance and
SU(3) symmetry. \\
~~~ These hyperons decay dominantly
into pions giving an additional contribution to the weak pion production induced by the
antineutrinos. In the case of nuclear targets like $^{40}$Ar, this contribution is shown to be significant when compared with the 
the pion production by the $\Delta$ excitations in the energy range of $E_{\bar{\nu}_{\mu}} \le 0.7$~GeV~\cite{Fatima:2018wsy}. This study could be useful for the DUNE experiment where argon will be used as the target material and the
electronic imaging of particles is possible and the particle tracks can be identified. 
 
\end{abstract}
%
\maketitle

\section{Introduction}
\label{intro}
The goal of the present day experimenters using accelerator neutrino and antineutrino beams at the ongoing experiments like T2K, NOvA, etc., or the planned experiments like DUNE at the Fermilab~\cite{fermi}, T2HK in J-PARC~\cite{JPARC}, is to determine the neutrino oscillation parameters with greater precision, and to search 
for the CP violation in the leptonic sector.

For this, the simultaneous knowledge of neutrino and antineutrino cross sections in 
the same energy region for a given nuclear target is required. Almost all the current generation neutrino experiments are using moderate to heavy nuclear targets, where the (anti)neutrino-nucleus cross section gets modified due to 
 the nuclear medium effects. In the few GeV energy region of neutrinos and antineutrinos, 
the contribution to the total scattering cross section comes from the quasielastic, inelastic and deep inelastic scattering of (anti)neutrinos with the nuclear targets. In this energy region, the inelastic channel is dominated by the single pion production, followed by the kaon production, associated particle production, multi-pion production, etc. In 
the case of antineutrinos, apart from the aforementioned processes, the quasielastic hyperon~($\Lambda$, $\Sigma^0$ and $\Sigma^-$) production  also contributes to the total scattering cross section. The produced hyperons then decay 
into a nucleon and a pion, thus, giving additional contribution to the single pion production. This mode of pion production is important in the low energy 
region, even though, the dominant contribution to the single pion production comes from the $\Delta$ excitations.
 We have estimated that in the low energy region of $E_{{\bar\nu}_\mu}$ 
$\le$ 0.7 GeV, the pions arising from the hyperon decays are comparable to the pions arising from the $\Delta$ excitations~\cite{Fatima:2018wsy}.
 This is because the pion production arising from the $\Delta$ excitations is considerably suppressed due to the
nuclear medium effects (NME) and the final state interaction (FSI)
effects as the $\Delta$ properties like its mass and decay width are modified in the nuclear medium. 
Furthermore, these deltas decay instantly to pions, which travel through the nucleus and  undergo final state interactions with the residual nucleus resulting in pion absorption or charge exchange reactions.
 On the other hand, the nuclear medium effects on the hyperon production cross sections are quite small. Moreover, the final state interactions of the produced hyperons, due to the strong interactions, in the presence
of the nucleons in the nuclear medium leading to elastic and
charge exchange reactions like $\Sigma N \rightarrow \Lambda N$ and $\Lambda N \rightarrow \Sigma  N$ affects only the relative yield of various hyperons~($\Lambda$ and $\Sigma$). The total number of pions produced through the different $Y \rightarrow N\pi$ decay modes is, therefore, least affected by the $YN \rightarrow YN$ final state interaction effects. 
 In addition, these pions are expected to be less affected by the final state interaction with the 
residual nucleus. This is because the hyperon decay widths are highly suppressed
in the nuclear medium making them live longer enabling them to  travel
through most of the nuclear medium before they decay. 
Therefore, the cumulative effects of the lower threshold
energy for the hyperon production compared to the delta production, and the near absence of the FSI 
for the pions coming from the hyperon decay compensate for
the Cabibbo suppression as compared to the pions coming from
the $\Delta$ excitations. This makes the study of hyperon production processes important in the
context of oscillation experiments with antineutrinos in the
sub-GeV energy region.

The study of single hyperon production is important in its own right as it provides information about the nucleon-hyperon 
transition form factors at high $Q^2$, which is presently known only at low $Q^2$ from the study of the semileptonic decays of neutron and  
hyperons. The symmetries of the weak hadronic current like the T-invariance, G-invariance and SU(3) symmetry can also be 
tested, and some information on the physics beyond the standard model may also be obtained.  { In the past, various calculations on the charged current induced (anti)neutrino and electron scattering from the free nucleon and nuclear targets, have been done by many authors in which both  the production cross sections as well as the  polarization observables of muons and baryons were studied~\cite{Akbar:2016awk,Fatima:2018gjy,Akbar:2017qsf,Bilenky:2013fra,Alam:2014bya,Bilenky:2013iua,Graczyk:2017rti,Kuzmin:2003ji,Wu:2013kla,Mintz:2001jc,Mintz:2002cj,Mintz:2004eu}. With the advancement in the detection technology and the observation of the $\tau$ neutrino~($\nu_{\tau}$) experimentally, some 
calculations for the polarization observables of the $\tau$ leptons produced in the $\nu_{\tau}(\bar{\nu}_{\tau})$ induced reactions have also been performed in the weak processes~\cite{Kuzmin:2003ji,Fatima:2020pvv,Kuzmin:2004ke,Graczyk:2004vg,Hagiwara:2004gs,Graczyk:2004uy,Valverde:2006yi}.   Recently, Thorpe et al.~\cite{Thorpe:2020tym} have also 
studied the effect of second class currents, axial dipole mass and SU(3) symmetry violation on the total and differential 
scattering cross sections in the antineutrino induced single hyperon production. We have also studied the dependence of different vector and axial vector current form factors including the second class current form 
factors on the total and differential cross section and have tested the degree of time reversal and G-parity violations in the 
antineutrino as well as electron induced single hyperon production~\cite{Fatima:2018wsy,Fatima:2018tzs,Akbar:2016awk,Fatima:2018gjy,Akbar:2017qsf,Fatima:2020pvv}. In this work, we have extended our model~\cite{Fatima:2018tzs,Fatima:2018gjy} for the production cross section and polarization observables of the hyperons produced in the antineutrino and electron induced reactions by taking into account the effect of SU(3) symmetry breaking, following the formalism of Faessler {\it et al.}~\cite{Faessler:2008ix} in the determination of the vector and axial vector  form factors in the strangeness sector. }

We have studied the polarization components of the hyperons produced in the antineutrino induced reactions and the effect of the various weak form factors on these polarization observables. The longitudinal~($P_L$) and perpendicular~($P_P$) components of 
polarization lie in the plane while the transverse~($P_T$) component of polarization lies perpendicular to the reaction plane and is 
forbidden by T-invariance. The experimental measurement of the non-zero value of the transverse component of the 
polarized hyperon will directly show the violation of time reversal in the weak sector. { Recently developed liquid argon time-projection chamber (LArTPC) technology which is being used in the MicroBooNE and ICARUS
experiments at the Fermilab 
and planned to be used in the proposed DUNE~(The Deep Underground Neutrino Experiment) experiment~\cite{DUNE,Acciarri:2016ooe}, with an expected fiducial mass of about 40kT, holds very high promises as LArTPC provide excellent tracking and calorimetry, real time 
full imaging of events where the detectors would be capable of
  identifying and measuring precisely the neutrino events over a wide range of energies. It would be possible to study the hyperon~($\Lambda,\Sigma$)
  production and its polarization by observing the asymmetry in the angular distribution of the pions when the hyperons decay into $N\pi$, as LArTPC gives the 3-dimensional track of the interaction.}  Therefore, it is feasible to study the physics of T-violation in the leptonic sector by measuring the 
polarization components of the hyperons produced in the antineutrino reactions~\cite{Fatima:2018wsy,Fatima:2018tzs,Fatima:2020pvv}. 

 In this work, we have focused our attention especially on the $\Lambda$ production in the antineutrino--nucleon/nucleus reactions, which is the most feasible channel to study experimentally. 
In section-\ref{formalism}, we describe, in brief, the formalism for calculating
the cross sections for the single hyperon production and the delta production from the free nucleon target.  The expression for the polarization components {\it viz.}, $P_L$, $P_P$ and $P_T$ of the hyperon
 produced in the quasielastic reactions, are given in section-\ref{pol}.
 Section~\ref{NME} deals with the effect of the nuclear medium on the
$\Delta$ and the hyperon productions, and the final state
interactions of the hyperons in the nuclear medium. The final
state interactions of the produced pions are discussed in section-\ref{FSI}. In section-\ref{result}, we present our results and finally in
section-\ref{summary}, we summarize and conclude the findings.

\section{Formalism}\label{formalism}
\subsection{Hyperon production off the free nucleon}
\subsubsection{Matrix element and form factors}
The general expression of the differential scattering cross section corresponding to the process
\begin{eqnarray}\label{process3}
 \bar{\nu}_\mu (k) + N (p) &\rightarrow& \mu^+ (k^\prime) + Y (p^\prime); \qquad \quad N= n,p; \qquad \quad Y=\Lambda, \Sigma
\end{eqnarray}
in the rest frame of the initial nucleon, is written as:
 \begin{eqnarray}
 \label{crosv.eq}
 d\sigma&=&\frac{1}{(2\pi)^2}\frac{1}{4E_{\bar{\nu}_\mu} M}\delta^4(k+p-k^\prime-p^\prime) \frac{d^3k^\prime}
 {2E_{k^\prime}}  \frac{d^3p^\prime}{2E_{p^\prime}} \overline{\sum} \sum |{\cal{M}}|^2,
 \end{eqnarray}
 where the transition matrix element squared is expressed as:
\begin{equation}\label{matrix}
  \overline{\sum} \sum |{\cal{M}}|^2 = \frac{G_F^2 \sin^2 \theta_c}{2} \cal{L}_{\mu \nu} \cal{J}^{\mu \nu},
\end{equation}
with $G_{F}$ being the Fermi coupling constant and $\theta_{C}$ being the Cabibbo angle.
The leptonic ($\cal{L}_{\mu \nu}$) and the hadronic ($\cal{J}^{\mu \nu}$) tensors are given by
\begin{eqnarray}\label{L}
\cal{L}^{\mu \nu} &=& ~\mathrm{Tr}\left[\gamma^{\mu}(1 + \gamma_{5}) \Lambda(k) \gamma^{\nu}
(1 + \gamma_{5}) \Lambda(k')\right], \\ 
\label{J}
\cal{J}_{\mu \nu} &=& \frac{1}{2} \mathrm{Tr}\left[\Lambda({p^\prime}) J_{\mu}
  \Lambda({p}) \tilde{J}_{\nu} \right], \qquad \quad \tilde{J}_{\nu} =\gamma^0 J^{\dagger}_{\nu} \gamma^0.
\end{eqnarray} 
The leptonic current is given by
 \begin{equation}\label{l}
 l^\mu = \bar{u} (k^\prime) \gamma^\mu (1 + \gamma_5) u (k),
\end{equation}
and the hadronic current is defined as
\begin{eqnarray}\label{j}
 {{J}}_\mu &=&  \bar{u} (p^\prime)\left( \left[ \gamma_\mu f_1^{NY}(Q^2)+i\sigma_{\mu \nu} 
 \frac{q^\nu}{M+M^\prime} f_2^{NY}(Q^2) + \frac{2 ~q_\mu}{M+M^\prime} f_3^{NY}(Q^2) \right]  \right.\nonumber\\
 &-&\left.\left[ \gamma_\mu \gamma_5 g_1^{NY}(Q^2) + 
  i\sigma_{\mu \nu} \frac{q^\nu}{M+M^\prime} \gamma_5 g_2^{NY}(Q^2) + \frac{2 ~q_\mu} {M+M^\prime}\gamma_5 g_3^{NY}(Q^2) 
   \right]\right) u (p),~~~
\end{eqnarray}
 with $M$ and 
 $M^\prime$ being the masses of the initial nucleon and the final hyperon. $q_\mu (= k_\mu - k_\mu^\prime = 
 p_\mu^\prime -p_\mu)$ is the four momentum transfer with $Q^2 = - q^2,~Q^2 \ge 0$. $f_1^{NY}(Q^2)$, $f_2^{NY}(Q^2)$ 
 and $f_3^{NY}(Q^2)$ are the vector, weak magnetic and induced scalar form factors and $g_1^{NY}(Q^2)$,~$g_2^{NY}(Q^2)$ 
 and $g_3^{NY}(Q^2)$ are the axial vector, induced tensor (or weak electric) and induced pseudoscalar form factors, 
respectively.

 Using the above definitions, the $Q^2$ distribution is written as
\begin{equation}\label{dsig}
 \frac{d\sigma}{dQ^2}=\frac{G_F^2 ~sin^2 \theta_c}{16 \pi M^2 E_{\bar{\nu}_\mu}^2} N(Q^2),
\end{equation}
where the expression of $N(Q^2)$ is given in the Appendix-A of Ref.~\cite{Fatima:2018tzs}.

For the determination of the vector and axial vector $N-\Lambda$ transition form factors, we take the following considerations into 
account: 
\begin{enumerate} 
\item[a)] The assumption of the SU(3) symmetry~\cite{Cabibbo:2003cu} and the conserved vector current (CVC) hypothesis leads to $f_3^{N\Lambda} (Q^2) = 0$ and the two 
vector form factors \textit{viz.} $f^{N\Lambda}_1(Q^2)$ and $f^{N\Lambda}_2(Q^2)$ are determined in terms of the electromagnetic 
form factors of the nucleon, \textit{i.e.} $f_{1}^{N}(Q^{2})$ and $f_{2}^{N}(Q^{2}),~ N=(p,n)$ as
 \begin{eqnarray}
 \label{fplambda}
 f_{1,2}^{p \Lambda}(Q^2)&=& -\sqrt{\frac{3}{2}}~f_{1,2}^p(Q^2).
 \end{eqnarray}
The electromagnetic form factors are expressed in terms of 
 the Sachs' electric and magnetic form factors $G_E^{p,n} (Q^2)$ and $G_M^{p,n} (Q^2)$ of the nucleons, for which various parameterizations are available in the literature and in our 
numerical calculations, we have used the parameterization given by Bradford et al.~\cite{Bradford:2006yz}.
 
\item[b)]  Assuming the SU(3) symmetry, the axial vector form factors $g_{1,2}^{p\Lambda}(Q^2)$ are rewritten in terms of $g_{1,2}^{pn}(Q^2)$ and 
$x_{1,2}(Q^2)=\frac{F^A_{1,2}(Q^2)}{F^A_{1,2}(Q^2)+D^A_{1,2}(Q^2)}$ as
\begin{eqnarray} \label{gplam}
 g_{1,2}^{p \Lambda}(Q^2)&=& -\frac{1}{\sqrt{6}}(1+2x_{1,2}) g_{1,2}^{np} (Q^2).
\end{eqnarray}
 We further assume that $F^A_{1,2}(Q^2)$ and $D^A_{1,2}(Q^2)$ have the same $Q^2$ dependence, such that $x_{1,2}(Q^2)$ 
become constant given by $x_{1,2}(Q^2)=x_{1,2}=\frac{F^A_{1,2}(0)}{F^A_{1,2}(0)+D_{1,2}^A(0)}$. 

 For the axial vector form factor $g_{1}^{pn}(Q^2)$, a dipole parameterization has been used:
\begin{eqnarray}\label{g1}
 g_{1}^{pn}(Q^2)=g_{A}(0)\left(1+\frac{Q^2}{M_{A}^2}\right)^{-2},
\end{eqnarray}
where $M_A$ is the axial dipole mass and $g_A(0)$ is the axial charge. For the numerical calculations, we have used the 
world average value of $M_A=1.026$ GeV. $g_A(0)$ and $x_1$ are taken to be 1.2723 and 0.364, respectively, as 
determined from the experimental data on the $\beta-$decay of neutron and the semileptonic decay of hyperons. 

 $g_2^{pn} (Q^2)$ is taken to be of the dipole form, in analogy with $g_{1}^{np}(Q^2)$, i.e., 
\begin{eqnarray}\label{g2} 
 g_{2}^{pn}(Q^2)=g_{2}^{pn}(0)\left(1+\frac{Q^2}{M_{2}^2}\right)^{-2}.
 \end{eqnarray}
In the numerical calculations we 
have taken real as well as imaginary values of $g_{2}^{pn} (0)$, with $|g_2 (0)|$ varying in the range $0-2$~\cite{Fatima:2018tzs}.

\item [c)] The pseudoscalar form factor $g_3^{N\Lambda} (Q^2)$ is proportional to the lepton mass and the contribution is 
small in the case of muon antineutrino scattering. However, in the numerical calculations, we 
have taken the following expression given by Nambu~\cite{Nambu:1960xd} using the generalized Goldberger-Treiman relation~\cite{Goldberger:1958vp}, 
 \begin{eqnarray}\label{Nambu}
  g_3^{N\Lambda} (Q^2) = \frac{(M + M^\prime)^2}{2~(m_K^2 + Q^2)} g_1^{N\Lambda} (Q^2),
 \end{eqnarray}
where $m_K$ is the mass of the kaon.
\end{enumerate}

For the discussion of $N-\Sigma$ transition form factors, please see Refs.~\cite{Fatima:2018tzs,Fatima:2018gjy}.

\subsubsection{SU(3) symmetry breaking effects}
 We have included the effect of SU(3) symmetry breaking on the vector and axial vector form factors following Ref.~\cite{Faessler:2008ix}, where the main features of the model may be summarized as:
 \begin{itemize}
  \item [i)] At the leading order, there is no symmetry breaking effect for the vector form factor $f_{1}^{N\Lambda}(Q^2)$ because of the  Ademollo-Gatto theorem~\cite{Ademollo:1964sr}.
  
  \item [ii)] The weak magnetic~($f_{2}^{N\Lambda} (Q^2)$) and axial vector~($g_{1}^{N\Lambda} (Q^2)$) form factors transform as
  \begin{equation}
   {\cal F}^{N\Lambda} = -\sqrt{\frac{3}{2}} \left(F + \frac{D}{3} +\frac{1}{9} \left(H_{1} -2H_{2} -3H_{3} -6H_{4}\right) \right),
  \end{equation}
where $F$ and $D$ are the SU(3) symmetric couplings while $H_{i}; i=1-4$ are the SU(3) symmetry breaking couplings. The values of these parameters are given in Ref.~\cite{Faessler:2008ix}, and are here quoted for $f_{2}^{N\Lambda} (Q^2)$:
\begin{eqnarray}
 D&=&1.237, \qquad \quad F~=~0.563, \qquad \quad H_{1} ~=~ -0.246 \nonumber \\
 H_{2} &=& 0.096, \qquad \quad H_{3} ~=~ 0.021, \qquad \quad H_{4} ~=~ 0.030 \nonumber 
\end{eqnarray}
and for $g_{1}^{N\Lambda}(Q^2)$:
  \begin{eqnarray}
 D&=&0.7505, \qquad \quad F~=~0.5075, \qquad \quad H_{1} ~=~ -0.050 \nonumber \\
 H_{2} &=& 0.011, \qquad \quad H_{3} ~=~ -0.006, \qquad \quad H_{4} ~=~ 0.037 .\nonumber 
\end{eqnarray}
For the $Q^2$ dependence of these SU(3) breaking couplings, we assume the dipole parameterization of the form
\begin{equation}
 H_{i}^{V,A} = \frac{H_{i}}{\left(1+\frac{Q^2}{M_{V,A}^2}\right)^2}
\end{equation}
where $V,A$, respectively, represents the vector and axial vector form factor, $M_{V}=0.84$ GeV and $M_{A}=1.026$ GeV represent the corresponding vector and axial dipole masses.

  \item [iii)] In the case of the second class currents, we have not considered any SU(3) symmetry breaking effects. 
  
  \item [iv)] Since the pseudoscalar form factor is parameterized in terms of the axial vector form factor, therefore, it receives the SU(3) symmetry breaking effect from $g_{1}^{N\Lambda} (Q^2)$.
 \end{itemize}

 \subsection{Polarization of the hyperon} 
 \label{pol}
 Using the covariant density matrix formalism, the polarization 4-vector~($\xi^\tau$) of the hyperon produced in 
 the reaction given in Eq.~(\ref{process3}) is written as~\cite{Bilenky}:
 \begin{eqnarray}\label{polar4}
\xi^{\tau}&=&\left( g^{\tau\sigma}-\frac{p'^{\tau}p'^{\sigma}}{{M^\prime}^2}\right) \frac{  {\cal L}^{\alpha \beta}  
\mathrm{Tr}\left[\gamma_{\sigma}\gamma_{5}\Lambda(p')J_{\alpha} \Lambda(p)\tilde{J}_{\beta} \right]}
{ {\cal L}^{\alpha \beta} \mathrm{Tr}\left[\Lambda(p')J_{\alpha} \Lambda(p)\tilde{J}_{\beta} \right]}.
\end{eqnarray}
 \begin{figure}
 \begin{center}  
        \includegraphics[height=5cm,width=9cm]{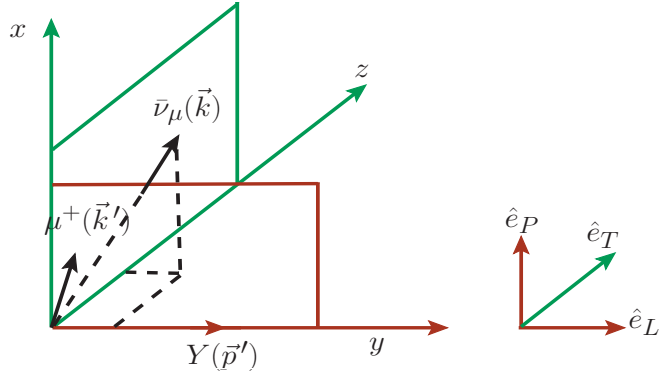}
  \caption{Polarization observables of the hyperon. $\hat{e}_{L}$, $\hat{e}_{P}$ and $\hat{e}_{T}$ represent the 
  orthogonal unit vectors corresponding to the longitudinal, perpendicular and transverse directions with respect to 
  the momentum of the final baryon.}\label{TRI}
   \end{center}
 \end{figure}
One may write the polarization vector $\bm{\xi}$ in terms of the three orthogonal vectors $\hat{e}_{i}~(i=L,P,T)$, 
i.e.
 \begin{equation}\label{polarLab}
\bm{\xi}=\xi_{L} \hat{e}_{L} + \xi_{P} \hat{e}_{P}+\xi_{T} \hat{e}_{T} ,
\end{equation}
where $\hat{e}_{L}$, $\hat{e}_{P}$ and $\hat{e}_{T}$ are chosen to be the set of orthogonal unit vectors corresponding 
to the longitudinal, perpendicular and transverse directions with respect to the momentum of the hyperon, depicted in 
Fig.~\ref{TRI}, and are written as
\begin{equation}\label{vectors}
\hat{ e}_{L}=\frac{\vec{ p}^{\, \prime}}{|\vec{ p}^{\, \prime}|},~~~~~
\hat{ e}_{P}=\hat{ e}_{L}\times \hat{ e}_T, ~~~~ 
\hat{e}_T=\frac{\vec{ p}^{\, \prime}\times \vec{ k}}{|\vec{ p}^{\, \prime}\times \vec{ k}|}.
 \end{equation}
The longitudinal, perpendicular and transverse components of the polarization vector $\bm{\xi}_{L,P,T} (Q^2)$ using 
Eqs.~(\ref{polarLab}) and (\ref{vectors}) may be written as:
\begin{equation}\label{PL}
 \xi_{L,P,T}(Q^2)=\bm{\xi} \cdot \hat{e}_{L,P,T}~.
\end{equation}
In the rest frame of the initial nucleon, the polarization vector $\bm{\xi}$ is expressed as
\begin{equation}\label{pol2}
 \bm{\xi} = A(Q^2)~ \vec{k} + B(Q^2)~ \vec{p}^{\, \prime} + C(Q^2)~  M (\vec{k} \times \vec{p}^{\,\prime})
\end{equation}
and is explicitly calculated using Eq.~(\ref{polar4}). The expressions for the coefficients $A(Q^2)$, $B(Q^2)$ and 
$C(Q^2)$ are given in the Appendix-A of Ref.~\cite{Fatima:2018tzs}.

The longitudinal ($P_L(Q^2)$), perpendicular ($P_P(Q^2)$) and transverse ($P_T(Q^2)$) components of the polarization 
vector in the rest frame of the final hyperon are obtained by performing a Lorentz boost and are written 
as~\cite{Fatima:2018tzs}:
\begin{equation}\label{PlPp}
 P_L (Q^2) = \frac{M^\prime}{E^\prime} \xi_L (Q^2), ~~~~~~~ P_P (Q^2) = \xi_P (Q^2), ~~~~~~~ P_T (Q^2) = \xi_T (Q^2).
\end{equation}
The expressions for $P_L (Q^2)$, $P_P (Q^2)$ and $P_T (Q^2)$ are then obtained using Eqs.~(\ref{vectors}), (\ref{PL}) 
and (\ref{pol2}) in Eq.~(\ref{PlPp}) and are expressed as
\begin{eqnarray}
P_L (Q^2) &=& \frac{M^\prime}{E^\prime} \frac{A(Q^2) \vec{k} \cdot \hat{p}^{\prime} + B (Q^2) |\vec{p}^{\,\prime}|}
  {N(Q^2)},  \label{Pl} \\
P_P (Q^2) &=& \frac{A(Q^2) [(\vec{k}.\hat{p}^{\prime})^2 - |\vec{k}|^2]}{N(Q^2) ~|\hat{p}^{\prime} \times \vec{k}|},
 \label{Pp} \\
 P_T (Q^2) &=& \frac{C(Q^2) M |\vec{p}^{\,\prime}|[(\vec{k}.\hat{p}^{\prime})^2 - |\vec{k}|^2]}{N(Q^2)~
 |\hat{p}^{\prime}  \times \vec{k}|}. \label{Pt}
\end{eqnarray}
To study 
the dependence of the polarization observables 
${P}_{L}(E_{\bar{\nu}_\mu})$, ${P}_P (E_{\bar{\nu}_\mu})$ and 
${P}_T (E_{\bar{\nu}_\mu})$ on  $E_{\bar{\nu}_\mu}$, we have integrated  
$P_L (Q^2),~P_P (Q^2)$ and $P_T (Q^2)$ over $Q^2$ and obtained the expressions for ${P}_{L,P,T} (E_{\bar{\nu}_\mu})$ as:
\begin{eqnarray}\label{average_Plpt}
 {P}_{L,P,T} (E_{\bar{\nu}_\mu}) &=& \frac{\int_{Q^2_{min}}^{Q^2_{max}} P_{L,P,T} (Q^2) 
 \frac{d\sigma}{dQ^2} dQ^2}{\int_{Q^2_{min}}^{Q^2_{max}} \frac{d\sigma}{dQ^2} dQ^2}.
\end{eqnarray}
If the T-invariance is assumed then all the vector and the axial vector form factors are real and the expression for 
$C(Q^2)$ vanishes which implies that the transverse component of polarization, $ P_T (Q^2)$ perpendicular to the 
production plane, vanishes.

\subsection{$\Delta$ production off the free nucleon}\label{deltafree}
 In the intermediate energy region of about $0.5 - 1$~GeV, the antineutrino induced reactions on a nucleon is 
 dominated by the $\Delta$ excitation, given by the reactions:
\begin{eqnarray}\label{eq1}
\bar\nu_\mu(k)+ n(p)&\rightarrow& \mu^{+}(k^\prime)+\Delta^{-}(p^\prime), \\
\label{eq2}
\bar\nu_\mu(k)+ p(p)&\rightarrow& \mu^{+}(k^\prime)+ \Delta^{0}(p^\prime),
\end{eqnarray}
 and the matrix element for the process given in Eq.~(\ref{eq1}) is written 
 as~\cite{Athar:2007wd}:
\begin{equation}\label{eq3}
T = \sqrt{3}\frac{G_F}{\sqrt{2}}\cos{\theta_c} ~l_{\mu} ~J^{\mu},
\end{equation}
where the leptonic current $l_\mu$ is defined in Eq.~(\ref{l}) and the hadronic current $J^\mu$ is given by
\begin{eqnarray}\label{eq4}
J^\mu&=&\overline{\psi}_\alpha(p^\prime)O^{\alpha\mu} u(p).
\end{eqnarray}
In the above expression, ${\psi_\alpha}(p^\prime)$ is the Rarita Schwinger spinor for $\Delta$ of momentum 
$p^\prime$ and $u(p)$ is the Dirac spinor for the nucleon of momentum $p$. $O^{\alpha\mu}$ is the $N-\Delta$ transition
operator which is the sum of the vector~($O^{\alpha\mu}_V$) and the axial vector~($O^{\alpha\mu}_A$) pieces, and the 
operators $O^{\alpha\mu}_V$ and $O^{\alpha\mu}_A$ are given by:
\begin{eqnarray}\label{eq5}
O^{\alpha\mu}_V&=&\left(\frac{C^V_{3}(q^2)}{M}(g^{\alpha\mu}{\not q}-q^\alpha{\gamma^\mu})+\frac{C^V_{4}(q^2)}{M^2}
(g^{\alpha\mu}q\cdot{p^\prime}-q^\alpha{p^{\prime\mu}})
\right. \nonumber \\
&+& \left.\frac{C^V_5(q^2)}{M^2}(g^{\alpha\mu}q\cdot p-q^\alpha{p^\mu}) + \frac{C^V_6(q^2)}{M^2}q^\alpha q^\mu\right)\gamma_5\\
\label{eq6}
O^{\alpha\mu}_A&=&\left(\frac{C^A_{3}(q^2)}{M}(g^{\alpha\mu}{\not q}-q^\alpha{\gamma^\mu})+\frac{C^A_{4}(q^2)}{M^2}
(g^{\alpha\mu}q\cdot{p^\prime}-q^\alpha{p^{\prime\mu}})+C^A_{5}(q^2)g^{\alpha\mu}\right.\nonumber\\
&+&\left.\frac{C^A_6(q^2)}{M^2}q^\alpha 
q^\mu\right).
\end{eqnarray}
A similar expression for $J^\mu$ is used for the $\Delta^0$ excitation without a factor of 
$\sqrt{3}$ in Eq.~(\ref{eq3}).  $C^V_i~(i =3-6)$ are the vector and $C^A_i(i=3-6)$ are the 
axial vector transition form factors, which are discussed in Ref.~\cite{paschos2}.

The differential scattering cross section for the reactions given in Eqs.~(\ref{eq1}) and (\ref{eq2}) is given 
by~\cite{Akbar:2017qsf,AlvarezRuso:1998hi,ruso}:
\begin{equation}\label{eq9}
\frac{d^2\sigma}{dE_{k^\prime}d\Omega_{k^\prime}}=\frac{1}{64\pi^3}\frac{1}{MM_\Delta}\frac{|{\vec{k}^\prime|}}
{E_k}\frac{\frac{\Gamma(W)}{2}}{(W-M_\Delta)^2+\frac{\Gamma^2(W)}{4}}{|{T}|^2},
\end{equation}
where $M_\Delta$ is the mass of $\Delta$ resonance, $\Gamma$ is the Delta decay width, $W$ is the center of mass energy 
i.e. $W=\sqrt{(p+q)^2}$ and
\begin{eqnarray}
 {|{T}|^2} &=& \frac{{G_F}^2\cos^2{\theta_c}}{2}L_{\mu\nu} J^{\mu\nu} \nonumber\\
 J^{\mu\nu} &=& \overline{\Sigma}\Sigma J^{\mu\dagger} J^\nu =\frac{1}{2} Tr\left[\frac{(\not p+ M)}{2 M}
 {\tilde{\mathcal O}}^{\alpha\mu } {\it P}_{\alpha \beta}{\mathcal O}^{\beta\nu} \right].
  \end{eqnarray}
In the above expression $L_{\mu\nu}$ is given by Eq.~(\ref{L}), 
 ${\tilde{\mathcal O}}^{\alpha\mu } = \gamma^0 {{\mathcal O}^{\alpha\mu }}^\dagger \gamma^0$, 
 ${\mathcal O}^{\alpha\mu }= O^{\alpha\mu}_V + O^{\alpha\mu}_A$ and $P^{\mu\nu}$ is the 
 spin $\frac{3}{2}$ projection operator defined as 
 \begin{equation}\label{prop}
  P^{\mu\nu}=
\sum_{spins}\psi^\mu {\overline{\psi}^\nu}=-\frac{\not{p^\prime}+M_\Delta}{2M_\Delta}\left(g^{\mu\nu}-\frac{2}{3}\frac{p^{\prime\mu} p^{\prime\nu}}
{M_\Delta^2}+\frac{1}{3}\frac{p^{\prime\mu} \gamma^\nu-p^{\prime\nu} \gamma_\mu}{M_\Delta}-\frac{1}{3}\gamma^\mu
\gamma^\nu\right).~~
 \end{equation}
In Eq.~(\ref{eq9}), the Delta decay width $\Gamma$ is taken to be an energy dependent P-wave decay width given 
by~\cite{a4}:
\begin{eqnarray}\label{gamma}
\Gamma(W)=\frac{1}{6\pi}\left(\frac{f_{\pi N\Delta}}{m_\pi}\right)^2\frac{M}{W}|{{\vec{q}}_{cm}|^3}\Theta(W-M-m_\pi),
\end{eqnarray}
where $f_{\pi N \Delta}$ is the $\pi N \Delta$ coupling constant taken as 2.12 for numerical calculations and 
$|{\vec{q}}_{cm}|$ is defined as
\[|{\vec{q}}_{cm}|=\frac{\sqrt{(W^2-m_\pi^2-M^2)^2-4m_\pi^2M^2}}{2W}.\]
The step function $\Theta$ in Eq.~(\ref{gamma}) denotes the fact that the width is zero for the invariant masses below 
the $N\pi$ threshold, and ${|\vec{q}_{cm}|}$ is the pion momentum in the rest frame of the $\Delta$ resonance.

\section{Nuclear medium effects}\label{NME}
\subsection{Hyperons produced inside the nucleus}\label{hyp-nu}
 When the antineutrino induced hyperon production  takes place on the nucleons which are bound in the nucleus, Fermi motion and Pauli 
 blocking effects of initial nucleons are considered. In the present work, the Fermi motion effects are calculated in a 
 local Fermi gas model~(LFGM), and the cross section is evaluated as a function of local Fermi momentum $p_F(r)$ and 
 integrated over the whole nucleus. The incoming antineutrino interacts with the nucleon moving inside the nucleus of 
 density $\rho_N(r)$ such that the differential scattering cross section inside the nucleus is expressed as 
 \begin{equation}\label{hyp-nucl}
\frac{d\sigma}{dQ^{2}}=2{\int d^3r \int 
\frac{d^3p}{{(2\pi)}^3}n_N(p,r)\left[\frac{d\sigma}{dQ^{2}}\right]_{\bar\nu N}},
\end{equation}
where $n_N(p,r)$ is the occupation number of the nucleon. $n_N(p,r)=1$ for $p\le p_{F_N} (r)$ and is equal to zero 
for $p>p_{F_N}(r)$, where $p_{F_N}(r)$ is the Fermi momentum of the nucleon and is given as:
$$
{p_F}_p(r) = \left(  3 \pi^2 \rho_p(r) \right)^\frac13;  \qquad \qquad {p_F}_n(r) = \left(  3 \pi^2 \rho_n(r) 
\right)^\frac13 ,
$$
with $\rho_p(r)$ and $\rho_n(r)$ are, respectively, the proton and the neutron densities inside the nucleus and are, 
in turn, expressed in terms of the nuclear density $\rho(r)$ as
\begin{eqnarray}
  \rho_{p}(r) &\rightarrow& \frac{Z}{A} \rho(r);  \qquad \qquad
  \rho_{n}(r) \rightarrow \frac{A-Z}{A} \rho(r). \nonumber 
 \end{eqnarray}
In the above expression, $\rho(r)$ is determined in the electron scattering experiments~\cite{vries}.

The produced hyperons are further affected by the FSI within the nucleus through the hyperon-nucleon elastic processes 
like $\Lambda N \rightarrow  \Lambda N$, $\Sigma N \rightarrow  \Sigma N$, etc. and the charge exchange scattering
processes like $\Lambda + n \rightarrow \Sigma^- + p$, $\Lambda + n \rightarrow \Sigma^0 + n$, $\Sigma^- + p 
\rightarrow \Lambda + n$, $\Sigma^- + p \rightarrow \Sigma^0 + n$, etc. Because of such types of interaction in the 
nucleus, the probability of $\Lambda$ or $\Sigma$ production changes and has been taken into account by using the 
prescription given in Ref.~\cite{Singh:2006xp}.

\subsection{Delta produced inside the nucleus}\label{del-nu}
 When an antineutrino interacts with a nucleon (Eq.\ref{eq1}) inside a nuclear target, nuclear medium effects like Fermi motion, Pauli blocking, etc., come 
 into play. The produced $\Delta$s have no such constraints in the production 
 channel but their decay is inhibited by the Pauli blocking of the final nucleons. Also, there are other disappearance 
 channels open for $\Delta$s through particle hole excitations and this leads to the modification in the mass and width 
 of the propagator defined in Eq.(\ref{gamma}).
 
 To take into account the nuclear medium effects, we have evaluated the cross section using the local density 
 approximation, following the same procedure as mentioned in section-\ref{hyp-nu}, and the differential scattering 
 cross section for the reactions given in Eqs.~(\ref{eq1}) and (\ref{eq2}) is defined as :
  \begin{eqnarray}\label{eq99}
\left. \frac{d^2\sigma}{dE_{k^\prime}d\Omega_{k^\prime}}\right|_{\bar\nu A}&=& \int d^3r \frac{1}{64\pi^3}\frac{1}{M
M_\Delta}\frac{|{\vec{k}^\prime|}} {E_k} \frac{\left(\frac{\tilde\Gamma(W)}{2} - Im\Sigma_\Delta\right)}{\left(W- 
M_\Delta - Re\Sigma_\Delta\right)^2 + \left(\frac{\tilde\Gamma(W)}{2} - Im\Sigma_\Delta \right)^2}\nonumber\\
&\times&\left(\rho_{n}(r)
~+~\frac{1}{3}\rho_{p}(r)\right) {|{T}|^2}.
 \end{eqnarray}
The modifications in the mass and decay width of $\Delta$ due to the nuclear medium effects, are given by
\begin{equation}
\frac{\Gamma}{2}\rightarrow\frac{\tilde\Gamma}{2} - Im\Sigma_\Delta, \qquad \quad
M_\Delta\rightarrow\tilde{M}_\Delta= M_\Delta + Re\Sigma_\Delta.
\end{equation}
The expressions of $Re\Sigma_\Delta$ and $Im\Sigma_\Delta$ are given in Ref.~\cite{a4}.

\section{Final state interaction effect}\label{FSI}
\subsection{Hyperon production}
 The hyperons decay to a pion and a nucleon {\it viz.}, $\Lambda \rightarrow p\pi^-, n\pi^0$, $\Sigma^{0} \rightarrow p\pi^-, n\pi^0$ and $\Sigma^-\rightarrow n\pi^-$. However, when the hyperons are produced in a 
 nuclear medium, some of them disappear through the hyperon-nucleon interaction processes like $YN \rightarrow NN$, 
 though it is suppressed due to the nuclear effects~\cite{Holstein,Oset:1989ey}. Moreover, the pionic decay modes of hyperons are Pauli 
 blocked as the momentum of the nucleons available in these decays is considerably below the Fermi level of energy for 
 most nuclei leading to a long lifetime for the hyperons in the nuclear medium~\cite{Holstein,Oset:1989ey}. Therefore, 
 the hyperons which survive the $YN \rightarrow NN$ transition in the medium live long enough to travel the nuclear medium 
 and decay outside the nucleus. In view of this, we have assumed no final state interaction of the produced pions with 
 the nucleons inside the nuclear medium. In a realistic situation, all the hyperons produced in these reactions will 
 not survive in the nucleus, and the pions coming from the decay of hyperons will undergo FSI~\cite{Singh:2006xp}. A 
 quantitative analysis of the hyperon disappearance through the $YN \rightarrow NN$ interaction and the pions having 
 FSI effect, will require a dynamic nuclear model to estimate the nonmesonic and mesonic decays of the hyperons in a 
 nucleus which is beyond the scope of the present work. Our results in the following section, therefore, represent an 
 upper limit on the production of pions arising due to the production of hyperons.
 
 \subsection{Delta production}\label{deltafsi}
The pions produced in the $\Delta$ excitation processes inside the nucleus may re-scatter or may produce more pions or may get absorbed 
while coming out from the final nucleus. We have taken the results of Vicente Vacas et al.~\cite{a5} for the final 
state interaction of pions which is calculated in an eikonal approximation using probabilities per unit length as the 
basic input. The details are given Ref.~\cite{a5}.

\section{Results and discussions}
\label{result}
\begin{figure}
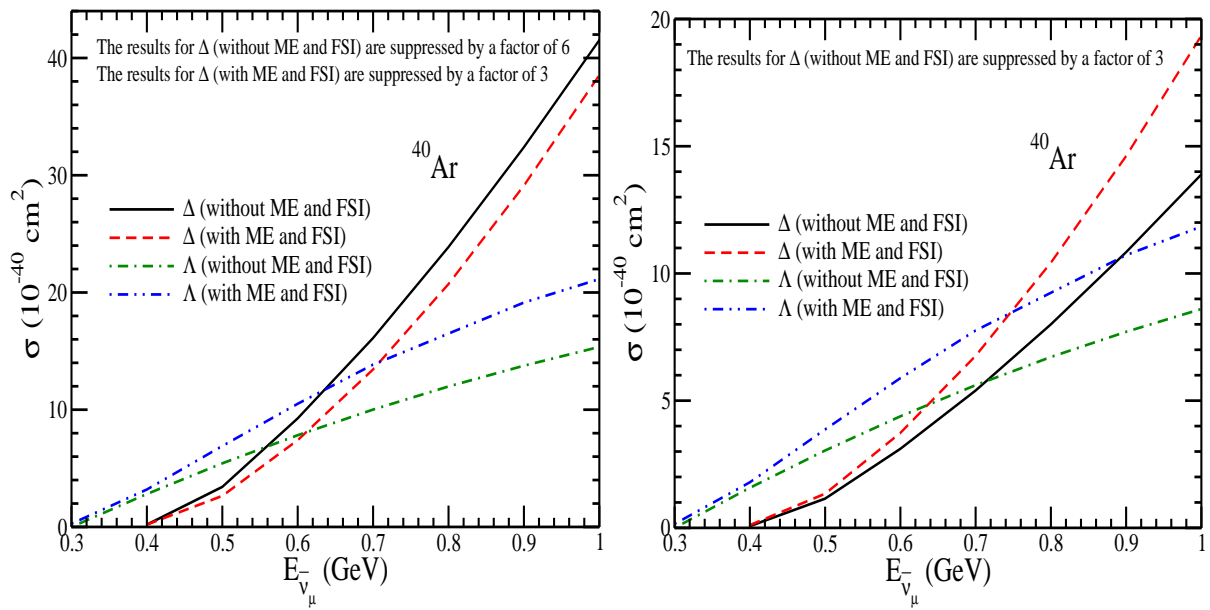

 \begin{center}
    \includegraphics[height=8cm,width=7.9cm]{total_sigma_argon_pi-_hyperon+delta_corrected.eps}
    \includegraphics[height=8cm,width=7.9cm]{total_sigma_argon_pi0_hyperon+delta_corrected.eps} 
       \end{center}
  \caption{Results for $\pi^-$~(left panel) and $\pi^o$~(right panel) production in  $^{40}$Ar with and without NME+FSI. The results are presented for the pion production from the $\Delta$ and $\Lambda$ with and without NME+FSI. Notice that in the case of $\pi^{-}$ production, the results of $\Delta$ without NME+FSI are suppressed by a factor of
  6 and the results for $\Delta$ with NME+FSI are suppressed by a factor of 3 to bring them on the same scale while in the case of $\pi^0$ production, the results of $\Delta$ without NME+FSI are suppressed by a factor of 3, and there is no scale factor when NME+FSI effects are included.}
   \label{fig:delta}
 \end{figure}

\subsection{Total scattering cross sections}
\label{subsec_total}
In Fig.~\ref{fig:delta}, the results for the total scattering cross sections $\sigma(E_{\bar\nu_\mu})$  vs $E_{{\bar\nu}_\mu}$ are presented  for ${\bar\nu}_\mu$ 
 scattering off $^{40}$Ar nuclear target,
 for the $\pi^-$(Left panel) and $\pi^o$(Right panel) 
productions, respectively. These results are shown for the cross sections obtained without and with nuclear medium and final state interaction (NME+FSI) effects 
for the pion production arising due to the $\Delta$ excitation and the $\Lambda$ production.  
Note the different scales used in Fig.~\ref{fig:delta} for displaying the results for $\pi^-$ and $\pi^0$ production cross sections as explained in the respective figure captions.
We have earlier discussed in some detail, the nuclear medium and final state interaction effects on the production of $\Delta$ and hyperon~($\Lambda + \Sigma$) 
in nuclei as well as the effect of nuclear absorption and charge exchange scattering of $\pi^0$ and $\pi^-$ in the nuclear medium~\cite{Fatima:2018wsy,SajjadAthar:2009rd}. We show, in Fig.~\ref{fig:delta}, the numerical results for $^{40}$Ar nucleus relevant in the low energy region of the DUNE experiment.
We find that in the case of pions produced through $\Delta$ excitations, NME+FSI leads to a 
 reduction of around $55 - 60\%$ in the $\pi^-$ production from the case when there is no NME+FSI,  for the antineutrino energies 0.6 $\le E_{\bar\nu_\mu}\le$ 
 1 GeV. Similarly, in the case of $\pi^0$ production from the $\Delta$ excitation, there is a reduction of about $50-60$\% when NME+FSI effects are included, in the energy range 0.6~GeV~$\le E_{\bar{\nu}_{\mu}} \le 1$~GeV. This reduction decreases with the increase in  antineutrino energies~\cite{SajjadAthar:2009rd}. In the case of $\pi^-$ production from $\Lambda$ decays, due to the inclusion of the FSI effect, there is a small enhancement at lower energies, which increases to $\sim 35\%$ at $E_{\bar{\nu}_{\mu}}=1$~GeV. However, in the presence of NME and FSI effects, the $\pi^-$ production from $\Delta$ decays dominates for $E_{\bar{\nu}_{\mu}}> 0.7$~GeV, below which $\pi^-$ from $\Lambda$ decays is quite comparable to the $\pi^-$ from $\Delta$ decays. Similarly, in the case of $\pi^0$ from the $\Lambda$ decays, the NME+FSI increases the production cross section by about 40\% at $E_{\bar{\nu}_{\mu}}=1$~GeV, and this enhancement increases with the increase in antineutrino energy. Moreover, it may be  noticed from the figure that the $\pi^0$ production from $\Lambda$ decays is comparable to $\pi^0$ from $\Delta$ excitations in the region from threshold up to $E_{\bar{\nu}_{\mu}}\le 0.75$~GeV, while at $E_{\bar{\nu}_{\mu}}<0.75$~GeV, $\pi^0$ from the $\Delta$ decays dominates.  It should be noted that the enhancement due to nuclear medium of $\pi^-$ and $\pi^0$ yield from the hyperon~($\Lambda + \Sigma$) decays is mainly due to $YN \rightarrow YN$ final state interactions  in which the $\Sigma$ hyperons get converted to $\Lambda$ hyperons.
 
  \begin{figure}
 \begin{center}
    \includegraphics[height=8cm,width=8cm]{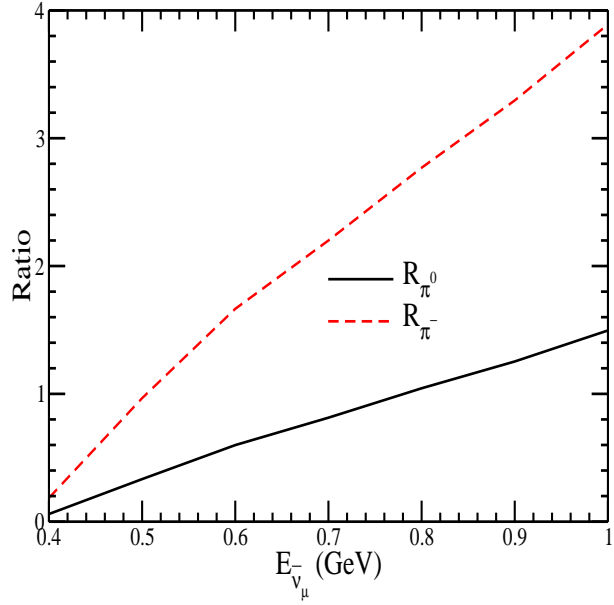} 
       \end{center}
  \caption{Results for the ratio $R_{\pi^0}$ ~(solid line) and $R_{\pi^-}$~(dashed line) in $^{40}$Ar  with NME+FSI.}
   \label{fig:ratio}
 \end{figure}
 
 \begin{figure}
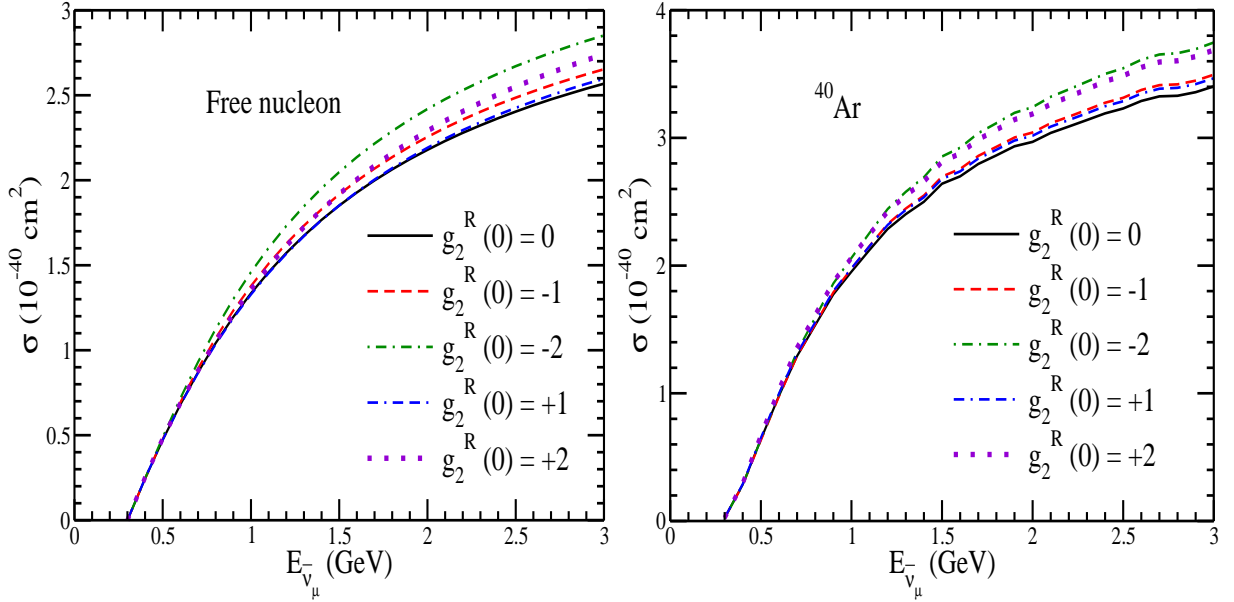

 \includegraphics[height=8cm,width=8cm]{sigma_free_g2R.eps}
 \includegraphics[height=8cm,width=8cm]{sigma_nucleus_g2R.eps}
\caption{$\sigma ~vs.~ E_{\bar{\nu}_{\mu}}$ for the process ${\bar{\nu}_\mu + p \rightarrow \mu^+ + \Lambda}$ for the free case~(left panel) and for the proton bound inside an Argon nucleus~(right panel), at the different values of $g_2^R (0)$ viz. 
$g_2^R (0) = $ 0~(solid line), $-1$~(dashed line),  $-2$~(dashed-dotted line), $+1$~(double-dashed-dotted line) and $+2$~(dotted line).}\label{fig1}
\end{figure}

\begin{figure}
 \includegraphics[height=8cm,width=8cm]{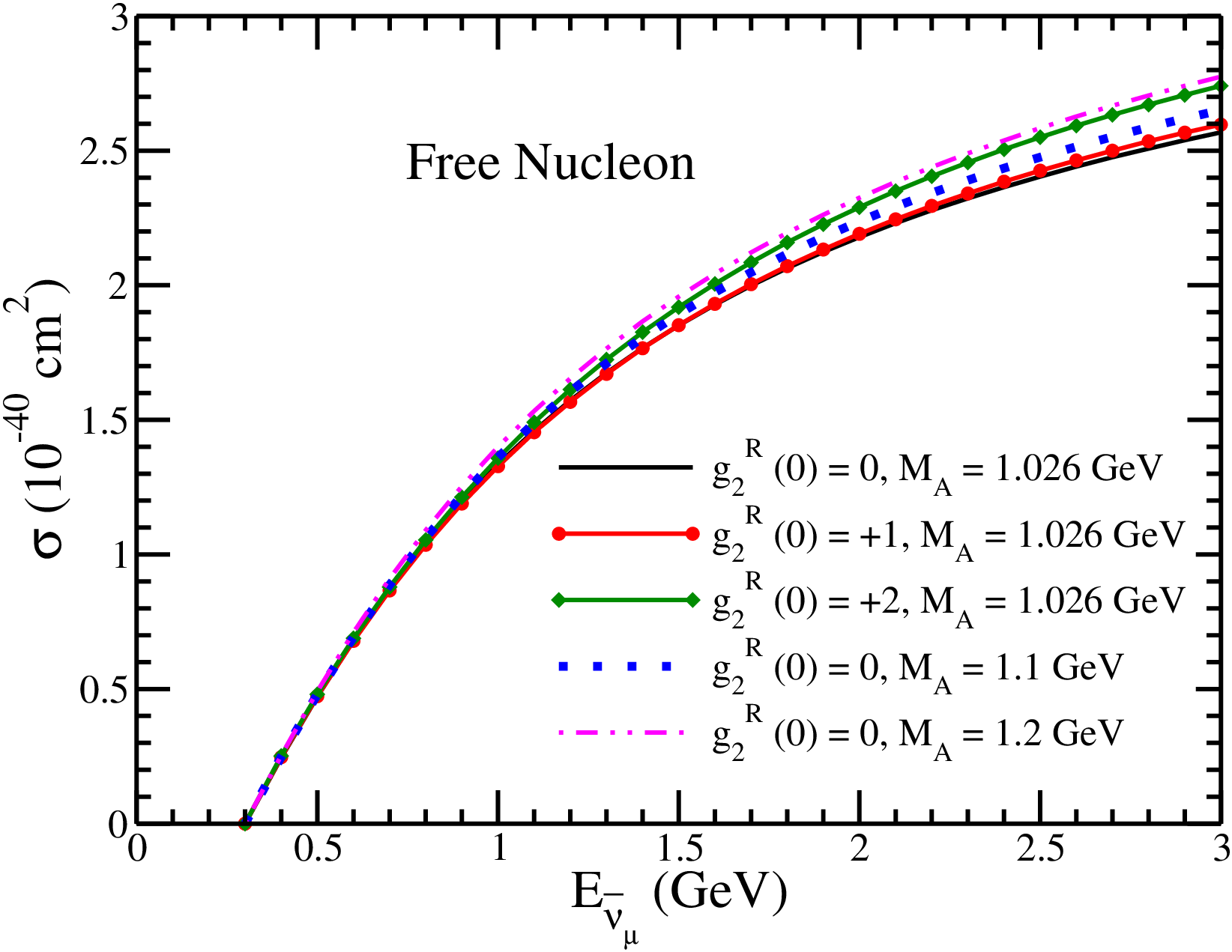}
 \includegraphics[height=8cm,width=8cm]{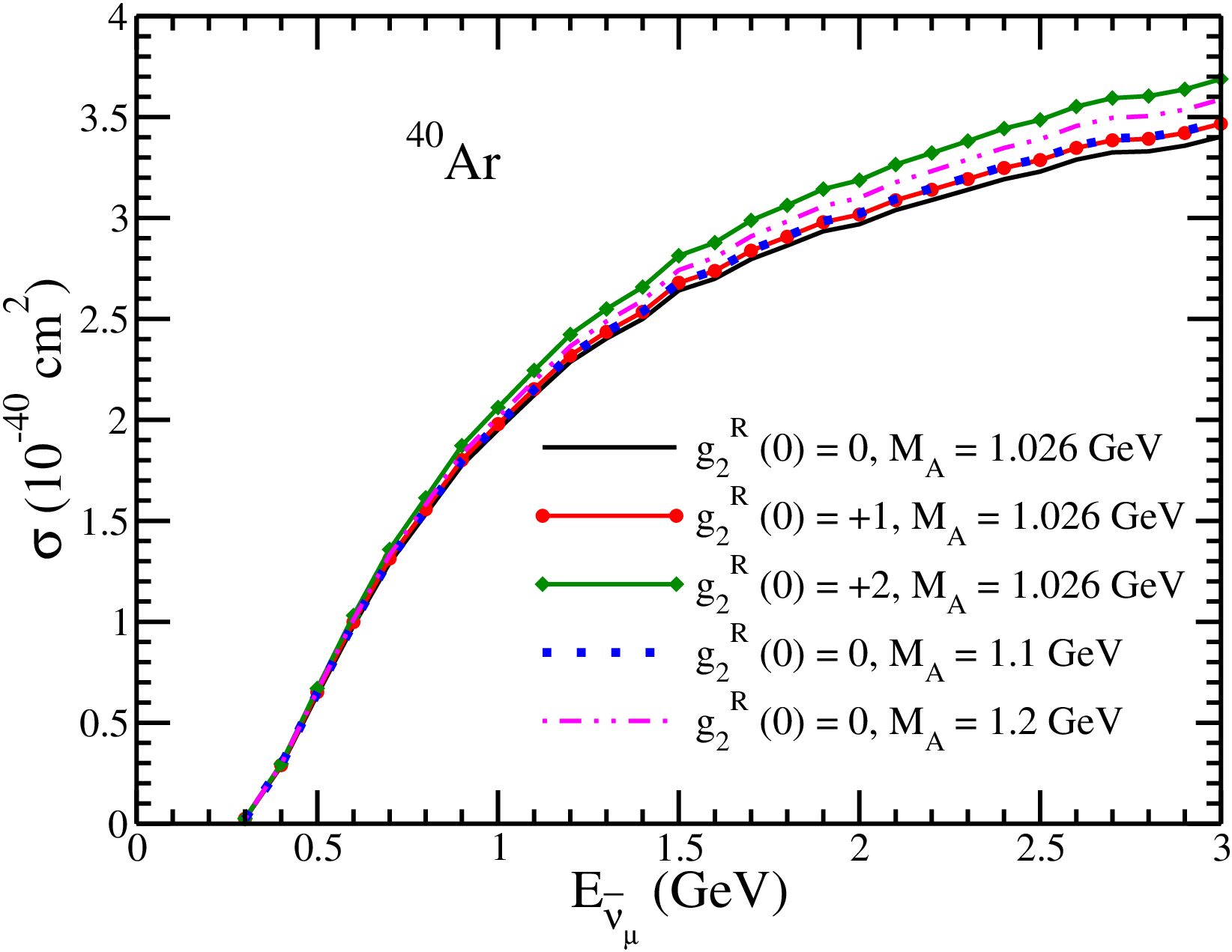}
\caption{$\sigma ~vs.~ E_{\bar{\nu}_{\mu}}$ for the process ${\bar{\nu}_\mu + p \rightarrow \mu^+ + \Lambda}$ for the free case~(left panel) and for the proton bound inside an Argon nucleus~(right panel), at the different values of $g_2^R (0)$ and the axial dipole mass, $M_{A}$, viz. 
$g_2^R (0) = 0$ and $M_{A}=1.026$ GeV~(solid line), $g_2^R (0) = +1$ and $M_{A}=1.026$ GeV~(solid line with circle), $g_2^R (0) = +2$ and $M_{A} = 1.026$ GeV~(solid line with diamond), $g_2^R (0) =0$ and $M_{A}=1.1$ GeV~(dotted line), and 
$g_2^R (0)=0$ and $M_{A} = 1.2$ GeV~(double-dotted-dashed line).}\label{fig2}
\end{figure}
 
\begin{figure}
 \includegraphics[height=8cm,width=8cm]{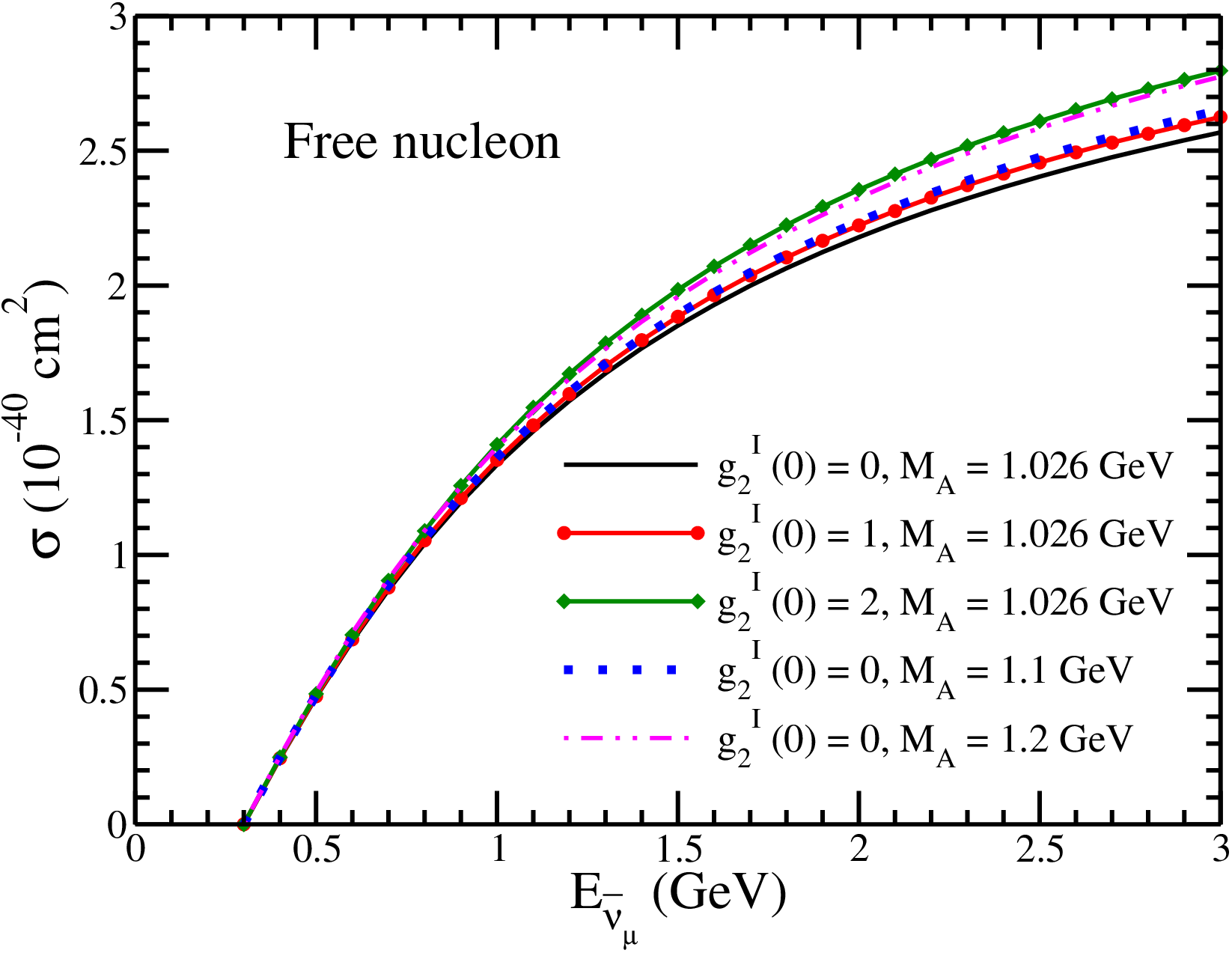}
 \includegraphics[height=8cm,width=8cm]{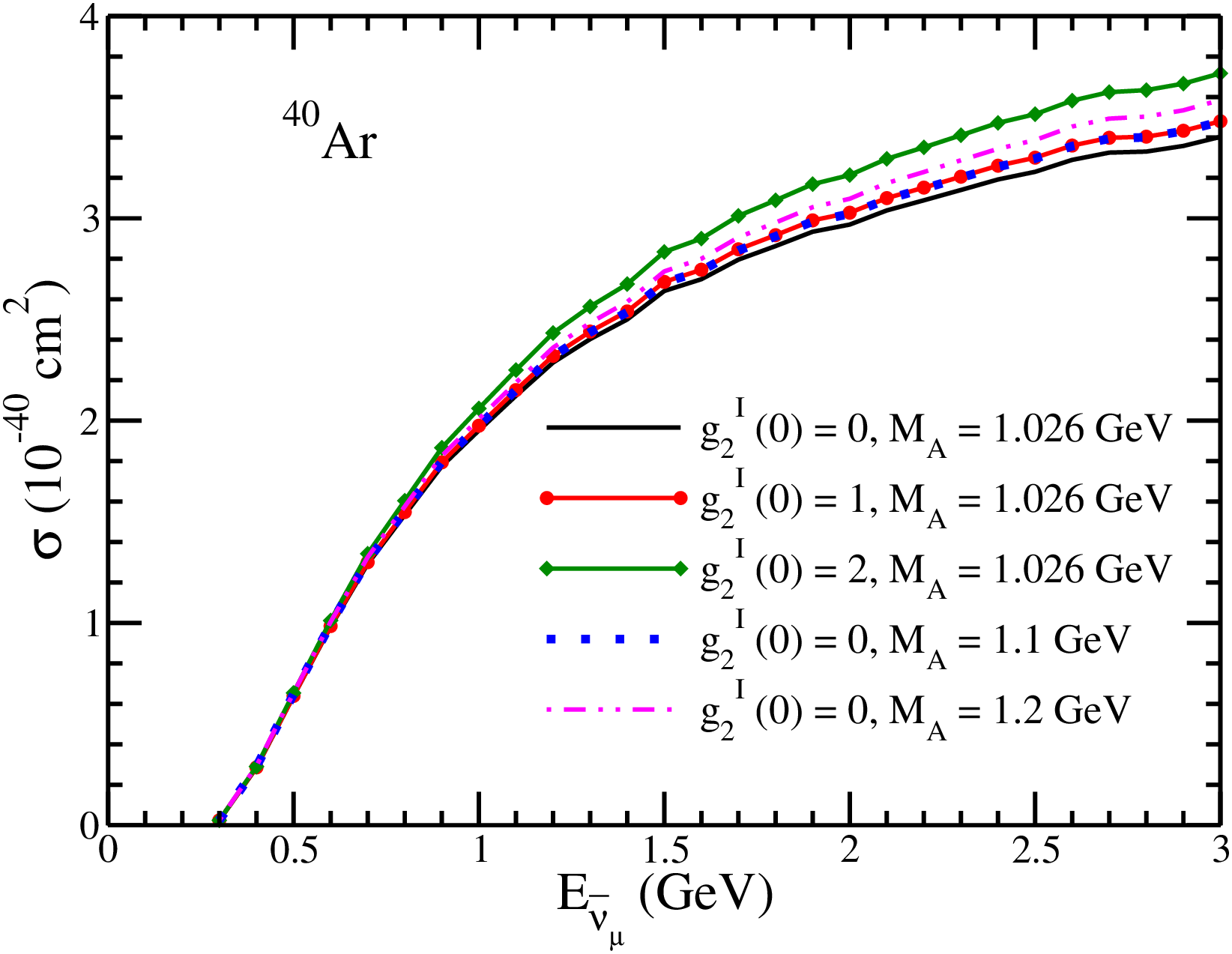}
\caption{$\sigma ~vs.~ E_{\bar{\nu}_{\mu}}$ for the process ${\bar{\nu}_\mu + p \rightarrow \mu^+ + \Lambda}$ for the free case~(left panel) and for the proton bound inside an Argon nucleus~(right panel), at the different values of $g_2^I (0)$ and the axial dipole mass, $M_{A}$, viz. 
$g_2^I (0) = 0$ and $M_{A}=1.026$ GeV~(solid line), $g_2^I (0) = 1$ and $M_{A}=1.026$ GeV~(solid line with circle),  $g_2^I (0) =2$ and $M_{A}=1.026$ GeV~(solid line with diamond), $g_2^I (0) = 0$ and $M_{A} = 1.1$ GeV~(dotted line), and $g_2^I (0) =0$ and $M_{A}=1.2$ GeV~(double-dotted-dashed line).}\label{fig3}
\end{figure}
 
To quantify the hyperon production cross sections and to compare the pions arising due to hyperons with the pions arising due to delta, 
in Fig.~\ref{fig:ratio}, we have obtained the results for the ratio
of delta to hyperon production cross sections, with NME+FSI, for $\pi^-$ and $\pi^o$ by defining
\begin{eqnarray*}
 R_{\pi^-} &=& \frac{\sigma(\Delta \rightarrow N\pi^-)}{\sigma(Y \rightarrow N\pi^-)}\\
 R_{\pi^o} &=& \frac{\sigma(\Delta \rightarrow N\pi^o)}{\sigma(Y \rightarrow N\pi^o)}
\end{eqnarray*}

It may be observed from these curves that in the case of $\pi^-$ production, at low energies, say around $E_{\bar{\nu}_{\mu}} \sim 0.5$~GeV, the ratio $R_{\pi^{-}}$ is approximately unity which implies that the number of $\pi^-$ coming from the hyperons  is almost comparable to the number of $\pi^-$  coming from $\Delta$. However, with the increase in $E_{\bar{\nu}_{\mu}}$, $\pi^-$ coming from $\Delta$ starts dominating and becomes $\sim$ 4 times the number of $\pi^-$ coming from the hyperons at $E_{\bar{\nu}_{\mu}}=1$~GeV.  In the case of $\pi^0$ production, as observed from Fig.~\ref{fig:ratio}, the ratio $R_{\pi^0}$ is less than one almost up to $E_{\bar{\nu}_{\mu}}\le 0.8$~GeV, which means that the $\pi^0$ arising from the hyperons has a larger contribution than the $\pi^0$ coming from $\Delta$, in the low energy region. However, with the increase in $E_{\bar{\nu}_{\mu}}$, the $\pi^0$ contribution from $\Delta$ starts dominating.    
 
 In Fig.~\ref{fig1}, we present the results for $\sigma(E_{\bar{\nu}_{\mu}}) ~vs.~ E_{\bar{\nu}_{\mu}}$ for the process ${\bar{\nu}_\mu + p \rightarrow \mu^+ + \Lambda}$ for the free nucleon~(left panel) and for the  argon nucleus~(right panel), at the different values of $g_2^R (0)$ viz. 
$g_2^R (0) = $ 0~, $\pm 1$,  $\pm 2$. In the case of the free nucleon target, we find that a non-zero value of $g_2^R (0)$ results in an increase in the cross section, which has been shown to be slightly larger for $g_2^R (0) < 0$ than for $g_2^R (0) > 0$ and this enhancement increases with the increase in the 
value of $g_2^R (0)$ and also with the increase in the energy of antineutrinos with a fixed $|g_{2}^{R} (0)|$. For example, with $g_2^R (0)=+1$ this enhancement from $g_2^R (0)=0$ is about $1-2\%$, while for $g_2^R (0)=-1$ the enhancement is about $3\%$, and with $g_2^R (0)=+2$ this enhancement from $g_2^R (0)=0$ is about $4\%$, while for $g_2^R (0)=-2$ the enhancement is about $8\%$ in the region of $E_{\nu_{\mu}} \sim 3$~GeV.
 While in the case of argon nucleus, we find that the difference in the numerical values of the cross sections calculated for the positive and negative values of $g_2^R (0)$, becomes negligible.  

Recently, in the literature there has been a lot of discussion on a larger value of the axial dipole mass $M_{A}$, compared to its world average value, in the quasielastic scattering
of (anti)neutrinos on the nuclear targets in order to explain the MiniBooNE, K2K and other results from the accelerator experiments.
A higher value of $M_A$, almost $10-20\%$ deviation from the world average value was suggested in the various analyses for explaining the results from the MiniBooNE experiment. For a detailed discussion, please see Refs.\cite{Alvarez-Ruso:2017oui,Katori:2016yel}. Therefore, to study whether a non-zero value of $g_2^R (0)$ has some effect on the total cross section for ${\bar{\nu}_\mu + p \rightarrow \mu^+ + \Lambda}$ scattering on the free nucleon~(left panel) and the nuclear target~(right panel), which can simulate the effect of using a higher value of $M_{A}$, we have taken a few non-zero values of $|g_2^R (0)|$ viz. 1 and 2 and  varied $M_A$ in the range 1.026 GeV to 1.2 GeV. It may be observed from Fig.~\ref{fig2}, that  $g_2^R (0)=2$
and $M_A=1.026$ GeV, simulate the cross section for $g_2^R (0)=0$
and $M_A=1.2$ GeV in the case of a free nucleon target. In the case of Argon nucleus, one may notice that the results obtained with $g_{2}^{R} (0) = 1$ and $M_{A}=1.026$~GeV simulate the results obtained with $g_{2}^{R}(0) = 0$ and $M_{A}=1.1$~GeV, while the results obtained with $g_{2}^{R}(0)=2$ and $M_{A}=1.026$~GeV are larger than the results obtained with $g_{2}^{R} (0)=0$ and $M_{A}=1.2$~GeV. However, the results with $g_{2}^{R} (0) = 1.5$ and $M_{A}=1.026$~GeV may simulate the results obtained with $g_{2}^{R} (0)=0$ and $M_{A}=1.2$~GeV. It suggests that a nonzero value of $g_{2}^R (0)$, can lead to a smaller value of $M_A$.

We have also considered T-violation and taken a non-zero imaginary value of $g_2(0)$, and 
 obtained the results for $\sigma ~vs.~ E_{\bar{\nu}_{\mu}}$, which are shown in Fig.~\ref{fig3}, for the process ${\bar{\nu}_\mu + p \rightarrow \mu^+ + \Lambda}$ for the free nucleon~(left panel) and for the argon nucleus~(right panel), with $M_A=1.026$~GeV, 1.1~GeV and 1.2~GeV with $g_2^{I}(0)=0,1$ and 2. It may be observed that the cross section not only increases with the increase in the value of $M_A$, but it also increases with the increase in the value of $g_2^I(0)$. The results of the cross section with $M_A=1.026$~GeV and $g_2^I (0)= 2$, simulate the results with $M_A=1.2$~GeV with $g_{2}^{I} (0) = 0$, i.e., when the second class current is switched off. It may be observed from the figure that this enhancement in the cross section with $M_A=1.026$~GeV and $g_2^I (0)= 2$, is more in the case of a nuclear target.

\begin{figure}
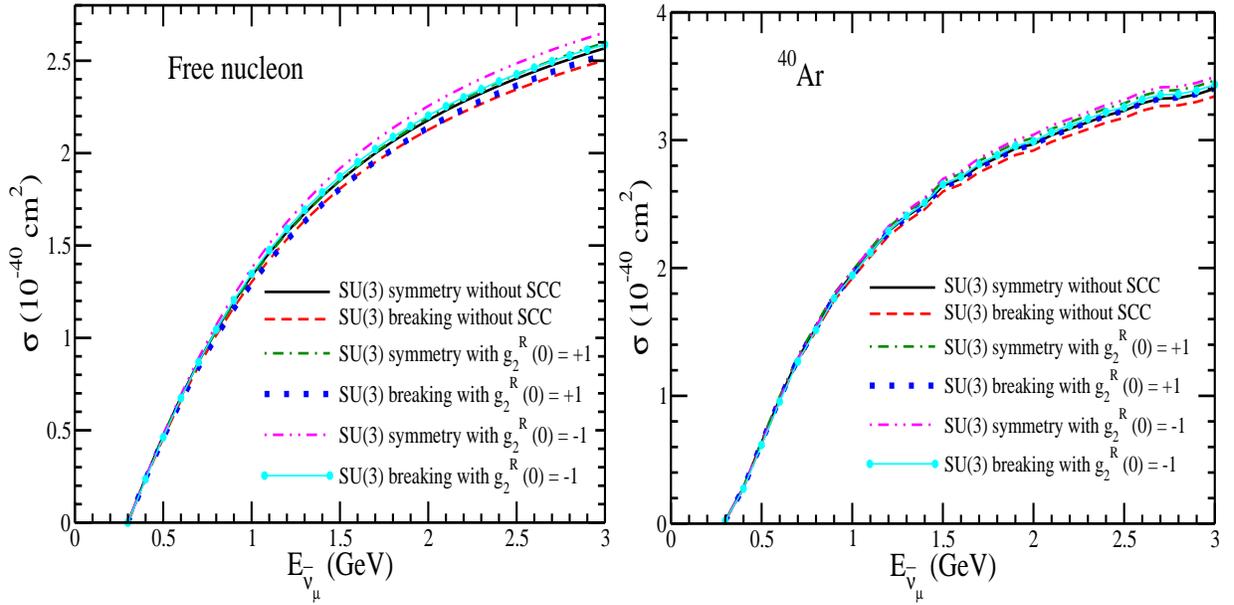

 \includegraphics[height=8cm,width=8cm]{SU3_breaking_real_g2_free_nucleon.eps}
 \includegraphics[height=8cm,width=8cm]{SU3_breaking_real_g2_argon.eps}
\caption{$\sigma ~vs.~ E_{\bar{\nu}_{\mu}}$ for the process ${\bar{\nu}_\mu + p \rightarrow \mu^+ + \Lambda}$ for the free case~(left panel) and for the proton bound inside an Argon nucleus~(right panel), with SU(3) symmetry breaking effect and in the presence of second class current form factor assuming T invariance.}\label{fig4}
\end{figure}

In Fig.~\ref{fig4}, we have presented the results for $\sigma ~vs.~ E_{\bar{\nu}_{\mu}}$ for the process ${\bar{\nu}_\mu + p \rightarrow \mu^+ + \Lambda}$ for the free nucleon~(left panel) and for the argon nucleus~(right panel), with SU(3) symmetry breaking effects and in the presence of second class current form factor assuming T invariance. We have already shown~(Fig.~\ref{fig1}) that the presence of the second class current form factor $g_{2}(Q^2)$ with or without T invariance increases the value of the cross section both for the free nucleon as well as for the argon nucleus. It may be observed from Fig.~\ref{fig4} that the effect of SU(3) symmetry breaking in the absence of the second class curent is to decrease the total cross section for the free nucleon as well as for the argon nucleus. Therefore, the two effects tend to cancel one another. In the case of the free nucleon, we observe that the results obtained with SU(3) symmetry breaking effect and $g_{2}^{R}(0) =-1$ is similar to the results obtained with SU(3) symmetry in the absence of the second class currents. In the case of argon nucleus, the results obtained with SU(3) symmetry breaking with $g_{2}^{R}(0)= + 1(-1)$ are similar to the results obtained assuming SU(3) symmetry in the absence of second class currents.

\begin{figure}
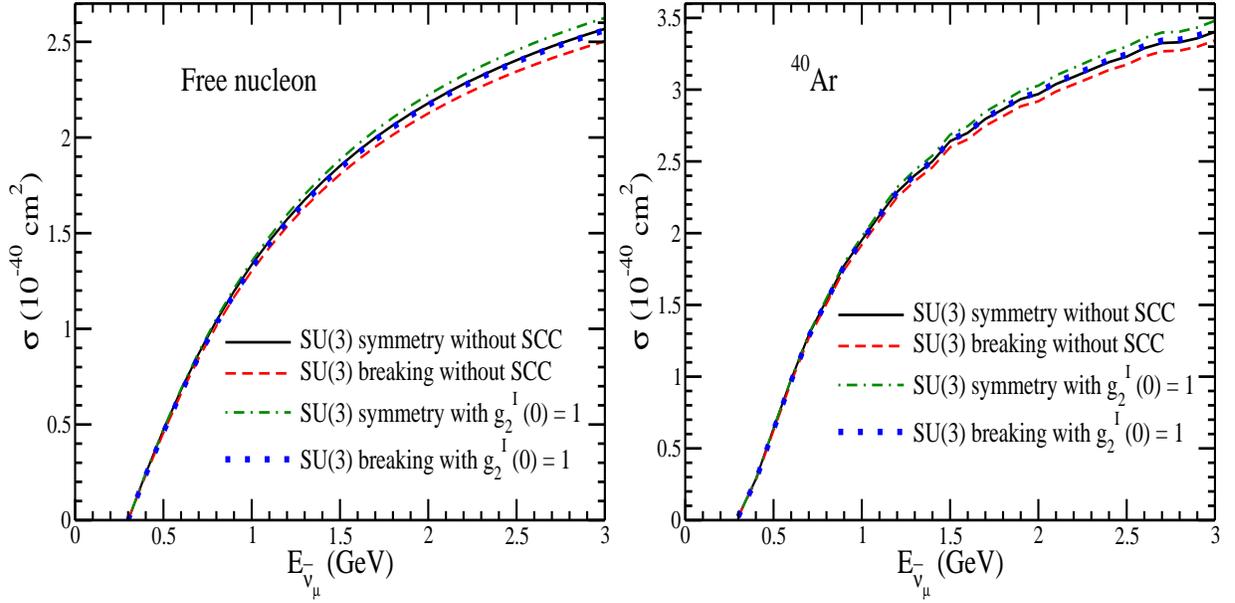

 \includegraphics[height=8cm,width=8cm]{SU3_breaking_imaginary_g2_free_nucleon.eps}
 \includegraphics[height=8cm,width=8cm]{SU3_breaking_imaginary_g2_argon.eps}
\caption{$\sigma ~vs.~ E_{\bar{\nu}_{\mu}}$ for the process ${\bar{\nu}_\mu + p \rightarrow \mu^+ + \Lambda}$ for the free case~(left panel) and for the proton bound inside an Argon nucleus~(right panel), with SU(3) symmetry breaking effect and in the presence of second class current form factor assuming T violation.}\label{fig5}
\end{figure}

In Fig.~\ref{fig5}, we have presented the results for $\sigma ~vs.~ E_{\bar{\nu}_{\mu}}$ for the process ${\bar{\nu}_\mu + p \rightarrow \mu^+ + \Lambda}$ for the free nucleon~(left panel) and for the argon nucleus~(right panel), with SU(3) symmetry breaking effects and in the presence of second class current form factor assuming T violation. In this figure, we observe similar results as in Fig.~\ref{fig4}. In the case of both free nucleon as well as the argon nucleus, the results obtained with  SU(3) symmetry breaking and $g_{2}^{I} (0) =1$ is similar to the results obtained with SU(3) symmetry in the absence of second class currents.

\subsection{Polarization observables}
\label{subsec_pol}
       \begin{figure}
 \begin{center}
      \includegraphics[height=8cm,width=8cm]{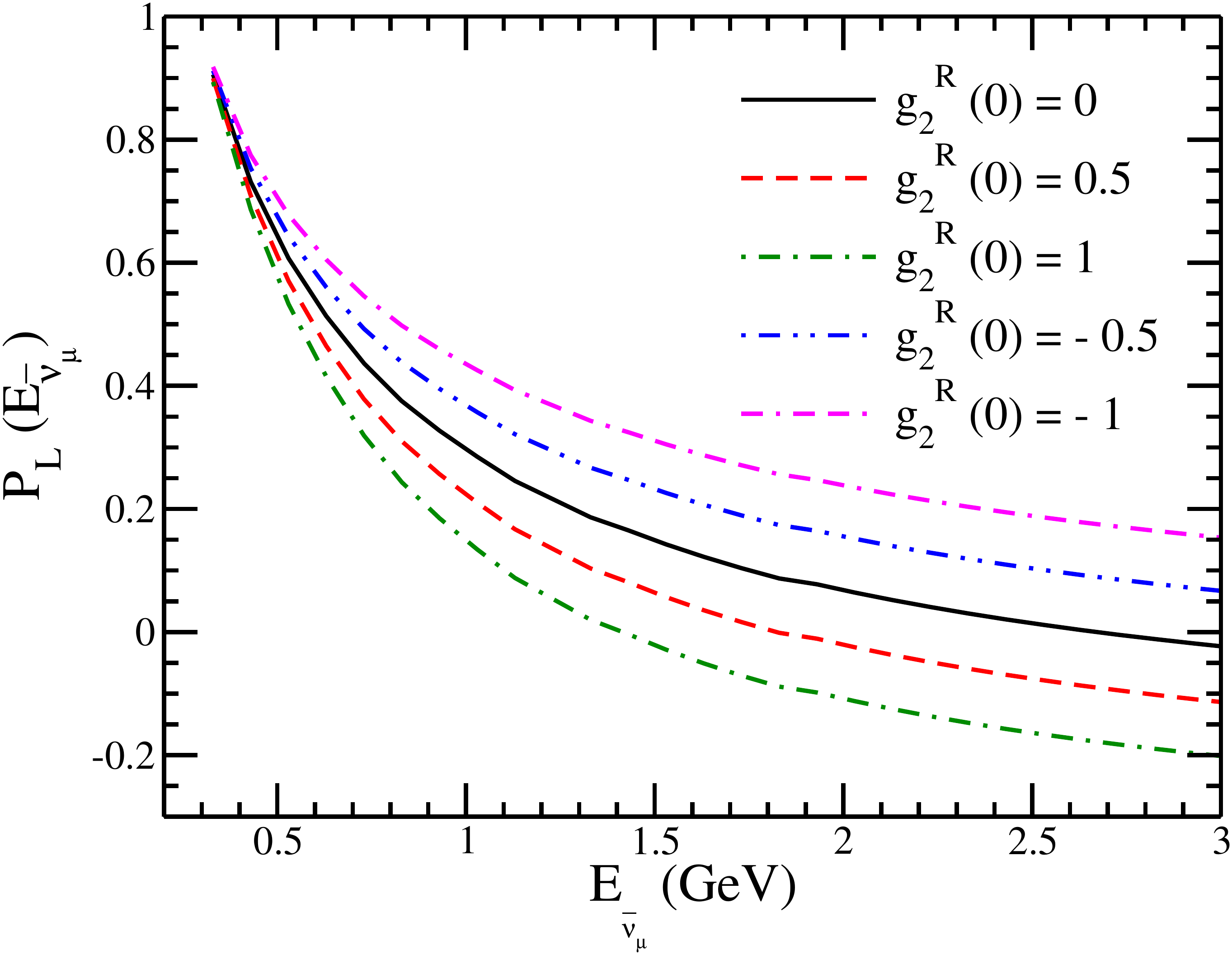}
 \includegraphics[height=8cm,width=8cm]{Pp_g2r_variation.eps}
       \end{center}
  \caption{$P_L (E_{\bar{\nu}_{\mu}})$~(left panel) and $P_P (E_{\bar{\nu}_{\mu}})$~(right panel) for the process $ \bar{\nu}_\mu + p \rightarrow \mu^+ 
  + \Lambda$ with $g_2^R(0) = $ 0~(solid 
  line), 0.5~(dashed line), 1~(dash-dotted line), -0.5~(double-dotted-dashed line) and -1~(double-dashed-dotted line).}
  \label{pol-2}
 \end{figure} 
 
 In Fig.~\ref{pol-2}, we have presented the results for the longitudinal~($P_{L}(E_{\bar{\nu}_{\mu}})$) and perpendicular~($P_{P}(E_{\bar{\nu}_{\mu}})$) components of the polarization for the  polarized $\Lambda$ produced in the reaction $\bar{\nu}_{\mu} +p \longrightarrow \mu^{+} + \Lambda$ and studied the effect of the nonvanishing real values of the weak electric form factor {\it viz.}, $g_{2}^{R}(0) = 0,~\pm 0.5$ and $\pm 1$ on the polarization observables. It may be observed from the figure that both $P_{L}(E_{\bar{\nu}_{\mu}})$ and $P_{P}(E_{\bar{\nu}_{\mu}})$ are quite sensitive to the variation of $g_{2}^{R} (0)$ for both positive as well as negative values. Qualitatively, for both the positive and negative values of $g_{2}^{R} (0)$, the longitudinal component of polarization shows similar trend while the perpendicular component of polarization shows different trends for positive and negative values of $g_{2}^{R} (0)$.
 
     \begin{figure}
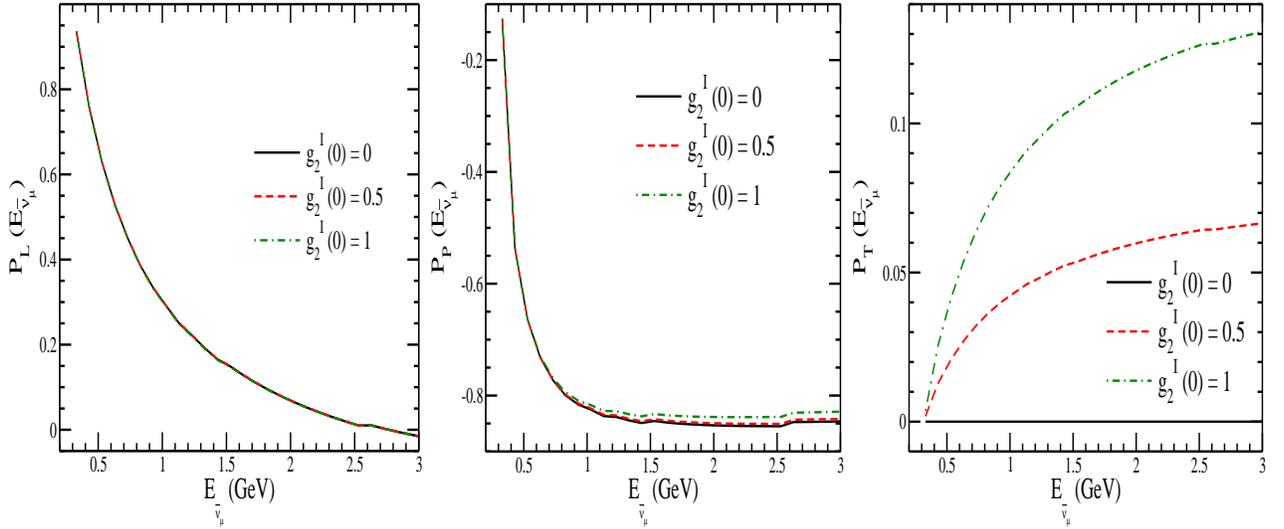

 \begin{center}
    \includegraphics[height=7cm,width=5.5cm]{Pl_g2I_variation.eps} 
    \includegraphics[height=7cm,width=5.5cm]{Pp_g2I_variation.eps} 
    \includegraphics[height=7cm,width=5.5cm]{Pt_g2I_variation.eps}
       \end{center}
  \caption{$P_L (E_{\bar{\nu}_{\mu}})$~(left panel), $P_P (E_{\bar{\nu}_{\mu}})$~(middle panel) and $P_{T} (E_{\bar{\nu}_{\mu}})$~(right panel) for the process $ \bar{\nu}_\mu + p \rightarrow \mu^+ 
  + \Lambda$ at different values of $g_{2}^{I} (0)$ {\it viz.}, $g_2^I(0) = $ 0~(solid 
  line), 0.5~(dashed line) and 1~(dashed-dotted line).}
  \label{pol-3}
 \end{figure} 
 
 In Fig.~\ref{pol-3}, we have presented the results for the longitudinal~($P_{L}(E_{\bar{\nu}_{\mu}})$), perpendicular~($P_{P}(E_{\bar{\nu}_{\mu}})$) and transverse~($P_{T}(E_{\bar{\nu}_{\mu}})$) components of the polarized $\Lambda$ produced in the reaction $\bar{\nu}_{\mu} +p \longrightarrow \mu^{+} + \Lambda$ and studied the effect of the non-zero imaginary values of the weak electric form factor {\it viz.}, $g_{2}^{I}(0) = 0, 0.5$ and 1 on the polarization observables. The results for $P_{L}(E_{\bar{\nu}_{\mu}})$ and $P_{P} (E_{\bar{\nu}_{\mu}})$ remain unchanged
when the negative values of $g_{2}^{I} (0)$ are taken as they depend
on $g_{2}^{R}(Q^2)$ and $|g_{2}^{R,I}(Q^2)|^{2}$ (see Appendix-A of Ref.~\cite{Fatima:2018tzs}). While in the case of $g_{2}^{I}<0$, $P_{T} (E_{\bar{\nu}_{\mu}})$ just changes sign but
the magnitude remains the same, and therefore, the results
have not been depicted in the figures. From the figure, it may be noted that the longitudinal as well as the  perpendicular components of polarization are insensitive to the variation in $g_{2}^{I} (0)$, but the transverse component of polarization, which appears only in the absence of T invariance, is quite sensitive to $g_{2}^{I} (0)$.  Even at $g_{2}^{I} = 0.5$, it could be around $5-7\%$ in the energy range of 1 to 3 GeV and increases with increase in the $g_{2}^{I} (0)$. 
         
\section{Summary and conclusions}
\label{summary}
In this review, we have studied the antineutrino induced quasielastic production of hyperons and their polarization components from the free nucleon as well as from the argon nucleus. These results may be useful
in determining the axial vector transition form factors in
the strangeness sector specially the form factor corresponding to the second
class currents with and without T-invariance. We have also studied the effect of SU(3) symmetry breaking on the cross section for hyperon production. 
\\

We summarize our results in the following:
\begin{itemize}
\item [i)] In the low energy region of antineutrinos, {\it i.e.}, $E_{\bar{\nu}_{\mu}} <0.7$~GeV, the pions produced by the decay of hyeprons~($\Lambda + \Sigma$) is comparable to the production of pions from the $\Delta$ decays. This is because the weak production of $\Delta$ in this energy region is inhibited by the threshold effect and is reduced by the nuclear medium effects. Moreover, the pions produced by the $\Delta$ decays undergo absorption and other final state interaction effects in the nuclear medium leading to a substantial reduction in the pion yield. These effects compensate for the Cabibbo suppression of the hyperon production as well as the reduction due to the nuclear medium and final state interactions in the case of the pions produced by the hyperon decays.

 \item [ii)] The total cross sections for the single hyperon production both from the free nucleon and from the argon nucleus are quite sensitive to the variation in the G violating axial vector form factor associated with the second class currents.
 
 \item [iii)] The non-zero value of the second class current form factor with and without T invariance may lead to a smaller value of the axial dipole mass in the case of $\nu_{\mu}(\bar{\nu}_{\mu})$ induced $\Delta S=0$ and $\Delta S=1$ quasielastic reactions from free as well as bound nucleons.
 
 \item [iv)] The presence of the second class current form factor with or without T invariance increases the total cross section with increase in $E_{\bar{\nu}_\mu}$, while the inclusion of the effect of the SU(3) symmetry breaking, in the case of $\Lambda$ production, decreases the cross section with the increase in $E_{\bar{\nu}_\mu}$, tending towards a cancellation due to these effects. 
 
 \item [v)] In the case of T invariance, the polarization components $P_{L} (E_{\bar{\nu}_{\mu}})$ and $P_{P} (E_{\bar{\nu}_{\mu}})$ are found to be sensitive to the variation in the value of $g_{2}^{R} (0)$, even for $g_{2}^{R} (0) = \pm 0.5$, and this sensitivity increases with the increase in $g_2^R(0)$.
 
 \item [vi)] In the case of T violation, the polarization components $P_{L} (E_{\bar{\nu}_{\mu}})$ and $P_{P} (E_{\bar{\nu}_{\mu}})$ do not show much sensitivity to the variation in $g_{2}^{I} (0)$, while the transverse component of polarization is found to be quite sensitive to $g_{2}^{I} (0)$ variation. This can provide an opportunity to study T violation by measuring the transverse polarization of $\Lambda$ in imaging detectors like the one planned to be used in DUNE.
\end{itemize}

\end{document}